# Critical thickness for ferromagnetism in LaAlO$_3$/SrTiO$_3$ heterostructures

Beena Kalisky[1-3,*], Julie A. Bert[1,2], Brannon B. Klopfer[1], Christopher Bell[2], Hiroki K. Sato[2,4], Masayuki Hosoda[2,4], Yasuyuki Hikita[2], Harold Y. Hwang[1,2] & Kathryn A. Moler[1,2]

1. Geballe Laboratory for Advanced Materials, Stanford University, Stanford, California 94305-4045, USA

2. Stanford Institute for Materials and Energy Sciences, SLAC National Accelerator Laboratory, Menlo Park, California 94025, USA

3. Department of Physics, Nano-magnetism Research Center, Institute of Nanotechnology and Advanced Materials, Bar-Ilan University, Ramat-Gan 52900, Israel

4. Department of Advanced Materials Science, University of Tokyo, Kashiwa, Chiba 277-8561, Japan

**In heterostructures of LaAlO$_3$ (LAO) and SrTiO$_3$ (STO), two nonmagnetic insulators, various forms of magnetism have been observed [1-7], which may [8, 9] or may not [10] arise from interface charge carriers that migrate from the LAO to the interface in an electronic reconstruction [11]. We image the magnetic landscape [5] in a series of n-type samples of varying LAO thickness. We find ferromagnetic patches that appear only above a critical thickness, similar to that for conductivity [12]. Consequently we conclude that an interface reconstruction is necessary for the formation of magnetism. We observe no change in ferromagnetism with gate voltage, and detect ferromagnetism in a non-conducting p-type sample, indicating that the carriers at the interface do not need to be itinerant to generate magnetism. The fact that the ferromagnetism appears in isolated patches whose density varies greatly between samples strongly suggests that disorder or local strain induce magnetism in a population of the interface carriers.**

For LAO grown on the TiO$_2$-terminated surface of the {100} face of STO, the interfaces are conducting [11] above a 3-unit-cell (3 uc) thickness [12] of LAO. In such interfaces, called "n-type", conductivity is attributed to interface electrons that are doped from the LAO into the conduction band of STO to solve the "polar catastrophe" of an electric potential that would otherwise diverge with the thickness of the LAO [11, 13]. However, the density of carriers inferred from Hall transport is much less than the charge density needed to resolve the polar catastrophe, and is also nearly an order of magnitude less than the spectroscopic charge density found in X-ray studies [14-16]. In addition, the interface is not conducting for any thickness of LAO grown on SrO termination layer (called "p-type") [11, 17], where an electronic reconstruction picture would predict that positively charged holes would migrate to the interface. These facts suggest that the polar catastrophe is an incomplete scenario: either competing atomic reconstructions occur to solve the potential divergence [17], or many of the interface charge carriers must be localized [14, 16] rather than mobile.

Magnetism at the interface has not been as extensively studied as conductivity. Magnetoresistance measurements [1-3] suggest local magnetic moments, and bulk magnetization measurements find a variety of behaviors [4]. In our previous work [5], we directly imaged micron-scale ferromagnetic patches with scanning Superconducting QUantum Interference Device (SQUID) microscopy. Polarized neutron reflectometry measurements [7] set an upper limit for surface magnetism that is at least 30 times smaller than that observed in some bulk measurements, but is still higher than the values we measured by scanning SQUID [5]. Theoretical work suggests that interface electrons may [8, 9] or may not [10] give rise to magnetism, and that the magnetism and should be highly sensitive to oxygen [18] and cation [19] disorder.

To investigate whether the interface electrons are essential to the magnetism, here we use scanning SQUID microscopy [20, 21] to magnetically image many samples (see Methods), primarily a series with varying LAO thickness. Figure 1a-d show representative examples of the magnetic landscapes for four thicknesses of the LAO layer. Our main observation is that a landscape of heterogeneous ferromagnetic patches similar to those shown in Fig. 1c and 1d appears in LAO/STO heterostructures starting at a LAO thickness of about 3 uc. Of eight samples with LAO layer thicknesses from 3 uc to 15 uc, seven have landscapes qualitatively similar to those in Fig. 1c and 1d, with strong sample-to-sample variability and heterogeneity, and two samples in this range (4 uc and 7 uc) show landscapes with a more dilute distribution of magnetic patches. The data on all three of the samples under 3 uc as well as four control samples (see Methods) are consistent with the absence of magnetism below the critical thickness.

To characterize the individual ferromagnetic patches, we fit each one to a model of a point dipole with five free parameters: the dipole moment's magnitude, azimuthal angle, inclination angle, and (x,y) position. In our measurement geometry (Fig. 1e and 1f), we scan the SQUID's pickup loop about a micron (a sixth fitting parameter) above the sample surface [21] while recording the total magnetic flux $\phi_s(x,y)$ through the pickup loop. The signal $\phi_s(x,y)$ is the local magnetic field of the sample integrated over the pickup loop area, with a sensitivity of 0.7 $\mu\phi_0/\sqrt{Hz}$ [21], where $\phi_0$ is one flux quantum, $\phi_0 = hc/2e$. An isolated micron- or submicron-scale ferromagnetic patch is conceptually similar to a small bar magnet with physical dimensions that are comparable to or smaller than the pickup loop. Figure 1g shows the flux $\phi_s(x,y)$ recorded while scanning the SQUID over an isolated magnetic patch. The size and intensity of the positive and negative lobes (green and red) depend strongly on the scan height as well as the characteristics of the patch itself, and the faint tails to the bottom of the dipole in the image are due to flux captured by the unshielded section of the leads to the pickup loop. Figure 1h shows a fitted image to compare to the data in Fig. 1g.

We can check whether the magnitudes of the observed dipole moments are consistent with a scenario of interface magnetism. The distribution of patch sizes and magnitudes does not vary systematically as a function of thickness above the critical thickness. In terms of spatial extent, most of the patches are both isolated and resolution-limited in size as compared to our 3 μm pickup loop scanning at a height of about 1 μm, so their maximal area could be as large as ~20 μm$^2$. Most of them may be much smaller, although in some cases there appear to be neighboring patches with different orientations or perhaps a much larger patch with internal domains. Under our measurement conditions, chosen to scan large areas on many samples, we can detect dipoles as small as ~10$^5$ Bohr magnetons ($\mu_B$) (flux signal of ~0.1 m$\phi_0$). The patches have a broad range of magnetic moments up to a few times 10$^8$ $\mu_B$ with a typical value around 10$^7$ $\mu_B$. Thus, we do not observe ferromagnetic patches that have a moment density per unit area much larger than ~1 $\mu_B$/uc at the interface. The distribution of sizes and magnitudes of the patches, as well as the fact that the maximum size does not change with LAO thickness, are consistent with a scenario of interface magnetism.

A ferromagnetic patch at the interface would be nearly two-dimensional, and would therefore have large shape anisotropies such that the magnetic moments would lie in plane. Indeed, the great majority of the magnetic moments do lie in-plane or nearly so. Crystalline anisotropy may also play a role. The azimuthal angle of the ferromagnetic patches appears nearly random, although in some samples there may be four-fold symmetry of the azimuthal orientation, possibly due to crystalline anisotropy. Due to this nearly random in-plane orientation, the vector sum of the moments in a typical data set such as in Fig. 1c and 1d would be much smaller than the sum of the magnitudes, making comparison to zero-field-cooled bulk measurements difficult. We note that the broad distribution of ferromagnetic patch sizes and shapes could give rise to odd behavior when making sample-averaged measurements under various sample histories and conditions, and may explain at least some of the apparent inconsistency in previous reports of magnetism.

To quantify the spatially heterogeneous distribution of ferromagnetism that we observe in all of the ferromagnetic samples, and to enable comparison to sample-averaged probes of magnetism, we calculate the average moment density in 50 μm x 50 μm regions as well as in each sample (Figure 2). For example, in the 5 uc sample (Fig. 2a), 16 out of the 24 regions shown in Fig. 2a had no detected magnetic moments larger than $2 \times 10^5 \mu_B$ (our sensitivity threshold). For comparison to other measurements and to theory, the highest density that we observe in any 50 μm x 50 μm region on any sample was ~$10^{12} \mu_B$/mm$^2$ ($10^{14} \mu_B$/cm$^2$), which, if associated with Ti atoms at the interface, would correspond to 0.15 $\mu_B$/Ti. Alternatively, if associated with the entire LAO layer in that sample, this density would correspond to 0.8 emu/cm$^3$ (taking 2 nm as the LAO layer thickness of a 5 uc sample) in a single 50 µm x 50 µm region. Most regions had much lower density, which would give an even smaller signal in bulk or sample-averaged measurements.

To compare the evolution of the conductivity and the magnetism, Fig. 2c and 2d show the average moment density (scalar) in each sample and the sheet conductance measured on the same samples as a function of LAO thickness. These data indicate a critical thickness of 3 uc for the appearance of ferromagnetic patches (the 2.7 uc sample is composed of 2 and 3 uc regions). In agreement with the literature, the heterointerface is insulating for LAO thicknesses less than or equal to 3 uc [12]. These observations are also consistent with a number of measurements demonstrating a precursor of the metal-insulator transition in the orbital reconstruction and the spectroscopic appearance of Ti$^{3+}$ occurring at 3 uc [16, 22], before macroscopic conductivity is observed above a percolation threshold [23].

To further probe the relationship between magnetism and conductivity, we tuned the conductivity of a 5 uc sample with a back gate [24, 25] (Fig. 3). This geometry allows for the application of a gate bias while preserving access to the sample surface by scanning SQUID. We found that the individual ferromagnetic patches do not change with gate voltage. Previous work showed that carrier concentration changes weakly with gating, while mobility changes rapidly [24]. We also looked for magnetism in non-conducting LAO/STO (p-type) [11, 17] (Fig. 4) and found that a p-type LAO/STO interface (STO terminated by a single unit cell of SrO) exhibits strong heterogeneous magnetism similar to the n-type samples. This indicates that the TiO$_2$ termination layer, which is essential for conductivity, is not necessary for magnetism.

To summarize, our main observation, that magnetism appears only above a threshold LAO thickness that is similar to the critical thickness observed in conductivity studies, supports a correlation between the appearance of magnetism and reconstruction of the interface. Other findings reported in this work (heterogeneity of ferromagnetism, sample to sample variability, magnetism in a non-conducting p-type interface, and independence of magnetism on gate voltage) all show that magnetism does not require the interface charge carriers to be mobile.

Theoretical work on the possibility of magnetism in LAO/STO can be approximately classified in two categories. In one category, the origin can be attributed to the band structure itself. Density functional theory (DFT) applied to a perfect interface [8] finds a number of nearly degenerate states, some of which have finite spin polarization on sites at and near the interface. In general, DFT predicts the occurrence of narrow, heavy electron mass sub-bands in the accumulation layer at the interface. Their propensity to localize in the presence of even small disorder has been suggested to account for the discrepancy between the measured Hall density and that found spectroscopically [9]. However, other theoretical work using dynamic mean-field theory does not suggest a stable magnetic ground state [10]. Whether localization could stabilize a spin-polarized state (including for p-type interfaces), and the microscopic origin of the disorder potential, remain to be clarified.

A second category of theoretical approaches identifies cation or oxygen vacancies as a source of local moments [19]. In particular, a recent model finds that incorporating oxygen vacancies at the LAO/STO interface [18] greatly enhances the tendency towards magnetism. While the results presented here cannot fully distinguish between these theoretical ideas, the importance of defects and disorder is clearly illustrated by both the sample-to-sample variability and by the heterogeneity within each sample, especially the existence of large regions (10's of microns) with little or no detectable ferromagnetism.

This heterogeneity far exceeds that which can be attributed to magnetic impurities or spatial variations in the cation stoichiometry. Thus oxygen, strain, and dislocations are all candidates for the relevant defect that tunes or seed the magnetism. The DFT work supports the importance of strain: if, as DFT suggests, there are moments at each site, the strain would change their separation [26] and therefore their coupling, possibly inducing ferromagnetism.

Nevertheless, the strong indications that the ferromagnetism reported here arises from the interface implies that it likely influences the superconductivity, which is known to exist at the interface below 300 mK [27]. The presence of both ferromagnetism and strong spin-orbit coupling at the interface may favor a topological superconducting phase that would support Majorana fermions [28, 29]. These results suggest the exciting possibility that the tuning from non-ferromagnetic to ferromagnetic may also tune the local superconductivity from conventional to exotic.

**Methods Summary**

$LaAlO_3$/$SrTiO_3$(100) samples were prepared by growing a certain number of unit cells (uc) of $LaAlO_3$ on commercial $TiO_2$ terminated {001} STO substrates. The growth process is as follows [24]: The $LaAlO_3$ was deposited at 800$^o$C with an oxygen partial pressure of $1.3 \cdot 10^{-5}$ mbar, after a pre-anneal at 950$^o$C with an oxygen partial pressure of $6.7 \cdot 10^{-6}$ mbar for 30 minutes. The samples were cooled to 600$^o$C and annealed in a high pressure oxygen environment (0.4 bar) for one hour.

The $LaAlO_3$ thicknesses measured in this study are nominally 1, 2, 2.7, 3.3, 4, 5, 7, 10 and 15 uc, as calibrated by *in-situ* reflection high-energy electron diffraction during growth. Another 10uc sample was grown on a SrO terminated $SrTiO_3$ to create the p-type interface.

Control samples measured in this study are: LaAlO$_3$ (100) as purchased, SrTiO$_3$ (100) as purchased, SrTiO$_3$ (100) annealed in the same conditions described for the thickness series above (defined as 0uc), SrTiO$_3$ annealed in a different chamber (post annealed in 1x10$^{-3}$ bar for 60 mins at 600$^o$C).

Although we tried to avoid magnetic contamination, small contamination while handling the samples is always possible [30] and could be responsible for the presence of an occasional dipole in the thinner samples: The low frequency of dipoles observed in these samples is consistent with the level of magnetism we find in other nonmagnetic samples and we attribute it to extrinsic contamination effects [30]. On two samples, one with 5 uc of LAO and one with 15 uc of LAO, we saw several larger dipole moments, in the range of 2 x 10$^8$ $\mu_B$ - 2 x 10$^9$ $\mu_B$. However, we are confident that these were not truly part of the LAO/STO heterostructure. To check whether they might be nanoparticles or dust contaminating the surface, we attempted to move them by scanning with the SQUID in contact with the sample and were successfully able to do so, unlike the vast majority of the smaller patches. To rule out contamination as a source of the magnetic signal we tested the local elemental composition with CAMECA NanoSIMS on two areas in the 15 uc sample. The Al and Ti used as references behaved as expected, and Cr, Mn, Ni, Co, Ru did not appear, with an upper limit of 1-10 ppm. The samples show some weak magnetic features, such as rare weak dipoles, and in some cases a weak magnetic landscape on a length scale of many tens of microns. These features are close to our resolution threshold, and are consistent with images we have seen on other nonmagnetic samples.

**Acknowledgments**

We thank Chuck Hitzman for his help with the NanoSIMS measurements, and Martin Huber for assistance in SQUID design and fabrication. We thank A.Millis, J.Mannhart, S.Raghu, and G.Sawatzky for helpful discussions. This work was supported by FENA-MARCO contract No. 0160SMB958, and partially supported by DARPA award No. C10J108234. This work was partially supported by the U.S. DOE, Division of Materials Sciences, under Award No. DE-AC02-76SF00515.

**Author Contributions**

Measurements: B.K. and J.A.B. Analysis: B.K. and B.B.K. with ideas developed with H.Y.H. and K.A.M. Sample growth: H.K.S., C.B., M.H, Y.H. and H.Y.H. Manuscript preparation: B.K and K.A.M., with input from all co-authors.

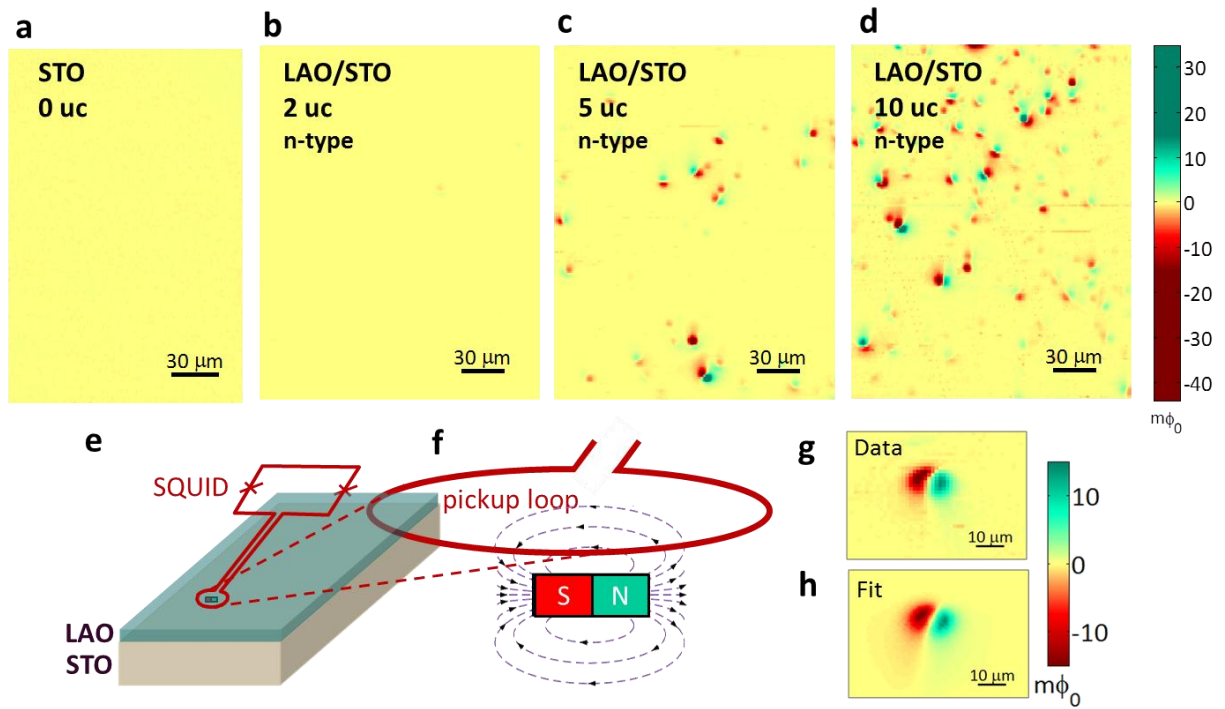

**Figure 1. Scanning SQUID images of the ferromagnetic landscape with increasing LAO thickness. (a-d)** Typical scanning SQUID images of the LAO/STO surface, showing no ferromagnetic patches for (a) annealed STO and (b) 2 uc (unit cell) of LAO, and ferromagnetic patches for (c) 5 uc and (d) 10 uc of LAO, taken at 4 K. **(e-f)** Sketch of magnetic imaging of a ferromagnetic patch with scanning SQUID. (e) Sketch of a SQUID probe with a 3 µm pickup loop near the surface of the sample. (f) Sketch of field lines from a ferromagnetic patch (conceptually shown here as a small bar magnet) captured in the pickup loop. (g) Image (data) of the flux through the pickup loop as the SQUID is scanned over a typical ferromagnetic patch. (h) Calculated image for an in-plane dipole whose moment is $7 \times 10^7$ Bohr magnetons and azimuthal angle $-20°$, as determined by fitting the data in (g) to a point dipole model.

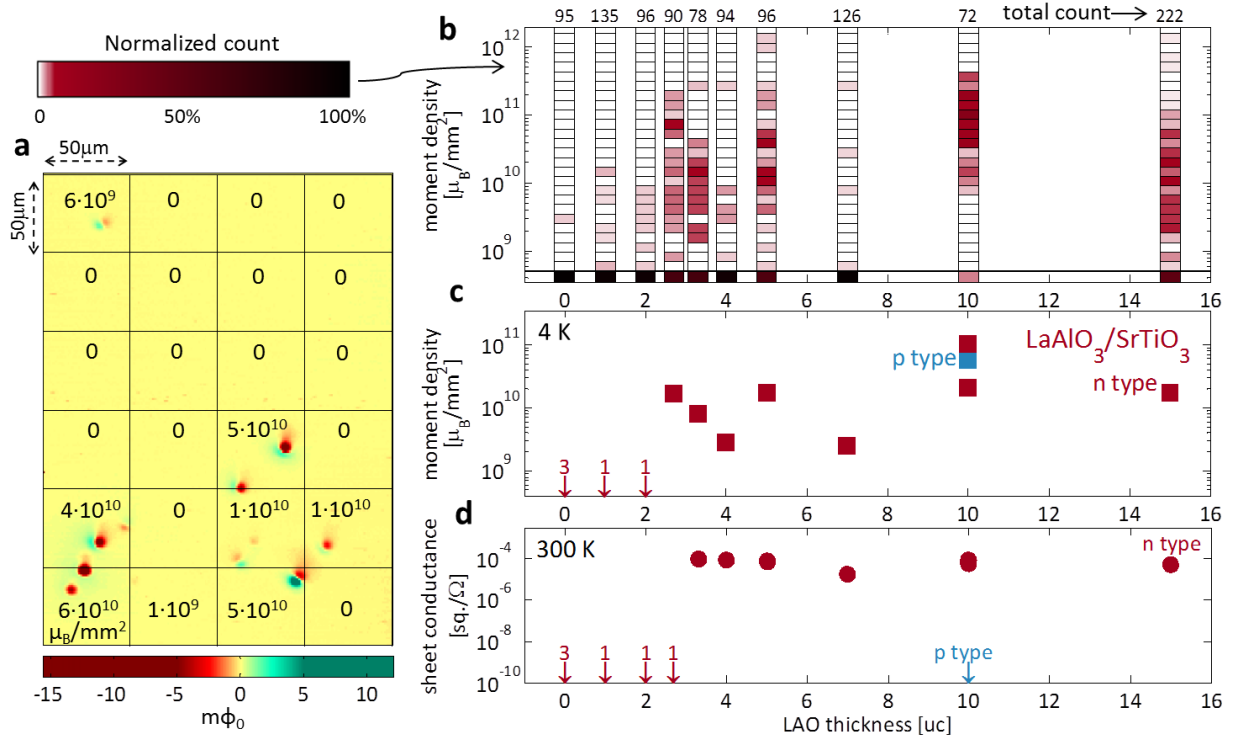

Figure 2. **Statistical distributions of magnetic moment density as a function of LAO thickness. (a)** Large area scan measured at 4 K in the 5 uc sample divided into 50 μm x 50 μm squares demonstrates the inhomogeneous distribution of the ferromagnetic patches. The numbers are the moment density in $\mu_B/mm^2$ for each square. 8 out of 24 squares have nonzero moment. **(b)** Histograms of the moment densities for ten samples of varying thickness. For each sample, colour indicates the percentage of squares that have a particular moment density. The number at the top of each column represents the number of 50 μm squares measured in each sample. For example, in the 15 uc sample we measured 222 squares of 50 μm x 50 μm and 75% of them had non-zero moment. Each histogram shows one representative sample per thickness. **(c)** Averaged moment density as a function of the thickness for TiO terminated (red) and SrO terminated (blue). Off scale values are marked by an arrow, which includes the number of samples for that thickness. **(d)** Sheet conductance at room temperature measured by the Van der Pauw method.

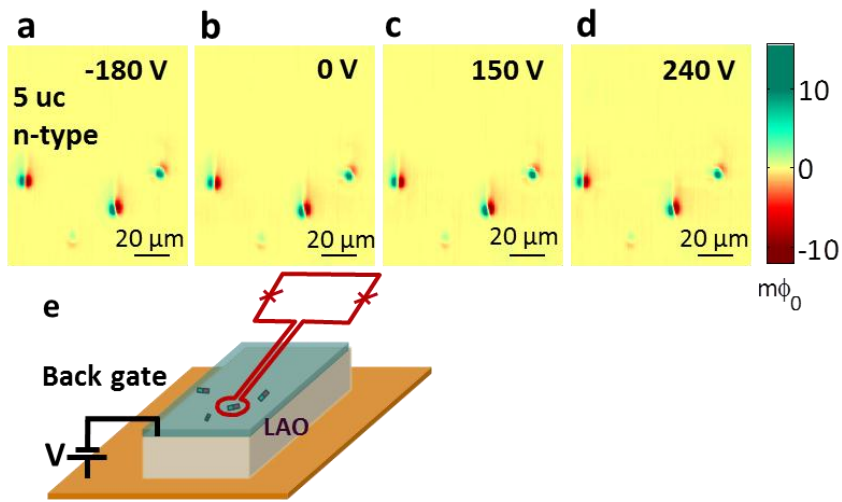

Figure 3. **The ferromagnetic patches do not change as a function of back gate voltage.** Scanning SQUID images repeatedly taken at 80 mK in a 5 uc n-type sample at back gate voltage of -180 V **(a)** 0 V **(b)** 150 V **(c)** and 240 V **(d)**. A sketch of the back gate configuration is shown in **(e)**.

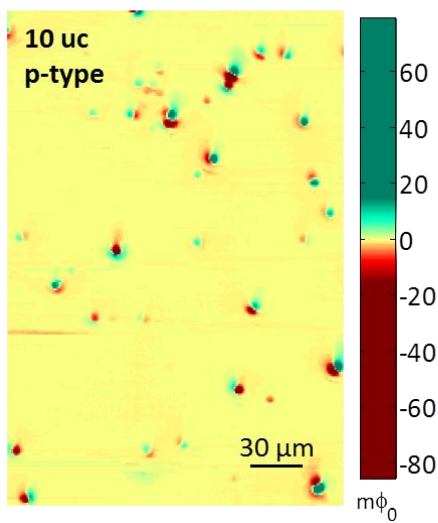

Figure 4. **Magnetic landscape in a p-type sample.** Scanning SQUID image shows a typical ferromagnetic landscape in a 10 uc non-conducting SrO terminated p-type LAO/STO sample at 4 K.

16